\documentclass[
 superscriptaddress,
 reprint,
 amsmath, amssymb,
 aps, preprintnumbers, nofootinbib, floatfix, longbibliography
]{revtex4-1}

\usepackage{graphicx}
\usepackage{dcolumn}
\usepackage{bm}
\usepackage{color}
\usepackage{amssymb}
\usepackage{amsmath}
\usepackage[colorlinks=true,allcolors=Purple,pdfusetitle]{hyperref}
\usepackage{rotating}
\usepackage{multirow}
\usepackage[dvipsnames]{xcolor}
\usepackage{svg}
\usepackage{physics}
\usepackage{comment}
\usepackage[T1]{fontenc}
\definecolor{Green}{RGB}{0, 128, 0}
\newcommand{\orcid}[1]{\href{https://orcid.org/#1}{\includegraphics[width=10pt]{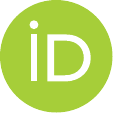}}}
\newcommand{\SN}{\vec{S}_N}
\newcommand{\Schi}{\vec{S}_\chi}
\newcommand{\qmN}{\frac{\vec{q}}{m_N}}
\newcommand{\Mtot}{M_{\text{tot}}}
\newcommand{\Erec}{E_{\text{rec}}}
\newcommand{\TeO}{TeO$_2$\,}
\newcommand{\vearth}{v_{\text{earth}}}
\newcommand{\Op}[1]{$\mathcal{O}_{#1}$}
\newcommand{\oxy}{$^{16}$O}

\begin{document}

\title{Uncertainties in tellurium-based dark matter searches stemming from nuclear structure uncertainties}

\author{Daniel~J.~Heimsoth~\orcid{0000-0002-5110-1704}}
\email{dheimsoth@wisc.edu}
\affiliation{
Department of Physics, University of Wisconsin--Madison,
Madison, Wisconsin 53706, USA}

\author{Rebecca~Kowalski~\orcid{0000-0001-6551-3948}}
\affiliation{
Department of Physics and Astronomy, The Johns Hopkins University, 
Baltimore, Maryland 21211, USA}

\author{Danielle~H.~Speller~\orcid{0000-0002-5745-2820}}
\affiliation{
Department of Physics and Astronomy, The Johns Hopkins University, 
Baltimore, Maryland 21211, USA}

\author{Calvin~W.~Johnson~\orcid{0000-0003-1059-7384}}
\email{cjohnson@sdsu.edu}
\affiliation{
Department of Physics, San Diego State University, 
San Diego, California 92182-1233, USA}

\author{A.~Baha~Balantekin~\orcid{0000-0002-2999-0111}}
\email{baha@physics.wisc.edu}
\affiliation{
Department of Physics, University of Wisconsin--Madison,
Madison, Wisconsin 53706, USA}

\author{Susan~N.~Coppersmith~\orcid{0000-0001-6181-9210}}
\affiliation{
School of Physics, The University of New South Wales,
Sydney, New South Wales 2052, Australia}

\date{January 2, 2025}
%update when resubmitting 

%%%%%%%%%%%%%%%%%%%%%%%%%%%%%%%%%%%%%%%%%%%%%%%%%%%%%%%%%%%%%%%%%%%%%%%%%%%%%%%%%%%%%%%%
%%%%%%%%%%%%%%%%%%%%%%%%%%%%%%%%%%%%%%%%%%%%%%%%%%%%%%%%%%%%%%%%%%%%%%%%%%%%%%%%%%%%%%%%

\begin{abstract}

Using tellurium dioxide as a target, we calculate uncertainties on 90\% upper confidence limits of Galilean effective field theory (Galilean EFT) couplings to a weakly-interacting massive particle (WIMP) dark matter candidate due to uncertainties in nuclear shell models. We find that these uncertainties in naturally-occurring tellurium isotopes are comparable across the different Galilean EFT couplings to uncertainties in xenon, with some reaching over 100\%. We also consider the effect these nuclear uncertainties have on estimates of the annual modulation of dark matter from these searches, finding that the uncertainties in the modulation amplitude are proportional to the non-modulating upper confidence limit uncertainties. We also show that the determination of the modulation phase is insensitive to changes in the nuclear model for a given isotope.

\end{abstract}

\maketitle

%%%%%%%%%%%%%%%%%%%%%%%%%%%%%%%%%%%%%%%%%%%%%%%%%%%%%%%%%%%%%%%%%%%%%%%%%%%%%%%%%%%%%%%%
%%%%%%%%%%%%%%%%%%%%%%%%%%%%%%%%%%%%%%%%%%%%%%%%%%%%%%%%%%%%%%%%%%%%%%%%%%%%%%%%%%%%%%%%

\section{Introduction}
\label{sec:Introduction}

Understanding the nature and form of dark matter (DM) has been one of the most pursued tasks in physics since the first concrete evidence of DM in galaxy cluster kinematics and galactic rotation curves\,\cite{Zwicky1933,Rubin1970,Freeman1970}. Many experiments have looked for microscopic evidence of particle DM interacting with Standard Model (SM) particles both directly and indirectly, with no unambiguous positive signal  thus far\,\cite{carlin2024cosine100dataset,COSINE-100_2022,darkside50_2023,LZ2024,XENONnT2024, pandax4t, cresstii2016, PhysRevD.99.062001, PhysRevD.100.022001, conradindir2017, heros2020, Roeck2024}. Many major hurdles exist, such as the apparently very small cross sections of DM-SM interactions, our lack of theoretical understanding of the form of the interaction, and the unknown characteristics of the DM particle.

A promising candidate for the DM particle is the Weakly Interacting Massive Particle (WIMP), which, as the name implies, has a nonzero mass and participates in forces no stronger than the weak interaction. In the past, interest in the WIMP was emboldened by the so-called ``WIMP miracle'' of supersymmetry, which predicted a stable, weakly-interacting particle with a mass of $100-1000$~GeV (though this range can be expanded by assuming different interaction coupling constants, see Ref.~\cite{Roszkowski:2017nbc} for a review of WIMPs). Even in searches for candidates from so-called ``light dark matter'' models probing the range below 10 GeV\,\cite{gelmini2017, kanekani1986, matsumoto2019, angevaare2022}, recent experiments have failed to find evidence for WIMPs. Such light dark matter has gained interest in direct detection searches in recent years as theoretical constraints relaxed\,\cite{Essig2012,XENONnT2024,XENONnT2024ionizationsignals,NEWS-G2024,NEON2024directsearchlightdark}. Until recently, analyses have mostly assumed either a simple spin-independent or spin-dependent DM-target interaction. Instead of making this assumption, a more general approach called Galilean effective field theory (Galilean EFT) considers all possible nonrelativistic, Galilean-invariant elastic interactions between DM and a target\,\cite{Fitzpatrick_2013} (for inelastic collisions, there is Chiral EFT; see Ref.~\cite{XENON1T_EFT2024}). Galilean EFT was developed at the level of nucleon fields; other effective theories start at quark and gluon fields and reduce to a bound, nonrelativistic state\,\cite{XENON1T_EFT2024,hoferichter2015chiral,hoferichter2019darkmatternucleus,Bishara_2017,Cirigliano:2012pq,Trickle:2020oki,hoferichter2019,Richardson2022}, though both approaches describe the most general interaction in the energy regime of interest for WIMP-nucleus scattering.

The experiments that could search for DM interactions directly (both those designed explicitly for DM direct detection and those with very low backgrounds, such as neutrinoless double beta decay experiments) generally choose large targets to maximize DM cross sections. For example, XENON1T and its successor XENONnT not surprisingly use xenon ($Z=54, N\approx75-80$), while CUORE uses the tellurium ($Z=52, N\approx70-78$) in \TeO crystals\,\cite{XENON1T_EFT2024,XENONnT2024,CUORE2016}. Because these nuclei contain many nucleons, they are themselves complicated quantum systems that must be approximated with models. For large nuclei, a popular approach is the configuration-interaction in a nuclear shell model basis, where ``valence'' nucleons fill valence energy and angular momentum states above a static core of nucleons, in analogy to atomic orbitals\,\cite{suhonen2007nucleons}. Calculating observable quantities in this basis, while reduced greatly from a full, infinite-dimensional configuration space, still consists of finding eigenvalues and eigenstates of a system with dimensions around and upward of tens of millions. These models have tunable parameters that can be fit to experimental data, allowing more freedom in building a particular model. The model approximation, parameter space truncation, and parameter tuning culminate in somewhat substantial theoretical uncertainties in predicted observables, cross sections being but one example.

Quantifying these nuclear modeling uncertainties is thus necessary to understand how they affect uncertainties of downstream calculations. Some of the current authors showed in Ref.~\cite{Heimsoth2023} how these nuclear modeling uncertainties lead to sizeable uncertainties in the determination of upper confidence limits of Galilean EFT coupling coefficients of WIMP-nucleus interactions for isotopes of xenon in the XENON1T detector. We continue that work in this paper by calculating these same uncertainties in Galilean EFT couplings for tellurium and oxygen as in the CUORE detector. Before providing the results of that analysis in Sec.~\ref{sec:uncerts}, we present an overview the theoretical framework of Galilean EFT and the DM-target interaction in Sec.~\ref{sec:theory}. In Sec.~\ref{sec:annualmod}, we consider the impact nuclear uncertainties have on determining parameters of a DM annual modulation signature. We end with a general discussion of the impact of this analysis in Sec.~\ref{sec:discussion} and conclusions and an outlook on future work in Sec.~\ref{sec:conclusion}.

\section{Theoretical Background}
\label{sec:theory}

\subsection{Nonrelativistic Effective Field Theory}

Historically, dark matter experiments have considered the simplest interactions between the DM particle and their target in their data analysis: a spin-independent density operator ($1_\chi 1_N$) and a spin-dependent operator ($\vec{S}_\chi \cdot \vec{S}_N$), where $\chi$ denotes the DM particle and $N$ denotes the target nucleus. However, these are only two of the many possible Hermitian interactions that obey Galilean invariance (or Lorentz invariance for relativistic theories). There has been a heightened interest recently in modeling DM-target interactions through effective field theories (EFT), which take a model-agnostic approach by considering all couplings allowed by imposed symmetries.

One such EFT is the Galilean EFT described in Ref.~\cite{Fitzpatrick_2013}, which provides CP-conserving coupling operators built from four Hermitian quantities:
$$i \vec{q},\quad \vec{v}^{\perp}, \quad \Schi, \quad \SN,$$
where $\Schi$($\SN$) is the spin of the DM (target) particle, $\vec{q}$ is the exchanged momentum, and $\vec{v}^{\perp} \equiv \vec{v} + \vec{q}/ 2 \mu_N$, $\vec{v}$ being the relative incoming velocity of the DM particle. Additionally, we use $\mu_N = m_\chi m_N / (m_\chi + m_N)$ to denote the reduced mass of the interaction, with $m_N$ the nucleon mass. The fifteen allowed operators $\mathcal{O}_i$ are
\begin{subequations}
\label{eq:operators} 
\begin{eqnarray}
	\mathcal{O}_1 &=& 1_\chi 1_N \ , \label{eq:O1} \\
	\mathcal{O}_2 &=& (v^\perp)^2 \ , \label{eq:O2} \\
	\mathcal{O}_3 &=& i \SN \cdot \left( \qmN \times \vec{v}^\perp \right) \ , \label{eq:O3} \\
	\mathcal{O}_4 &=& \Schi \cdot \SN \ , \label{eq:O4}\\
	\mathcal{O}_5 &=& i \Schi \cdot \left( \qmN \times \vec{v}^\perp \right) \ , \label{eq:O5} \\
	\mathcal{O}_6 &=& \left( \Schi \cdot \qmN \right) \left( \SN \cdot \qmN \right)  \ , \label{eq:O6} \\
	\mathcal{O}_7 &=& \SN \cdot \vec{v}^\perp \ , \label{eq:O7} \\
	\mathcal{O}_8 &=& \Schi \cdot \vec{v}^\perp \ , \label{eq:O8} \\
	\mathcal{O}_9 &=& i \Schi \cdot \left( \SN \times \qmN \right) \ , \label{eq:O9} \\
	\mathcal{O}_{10} &=& i\SN \cdot \qmN \ , \label{eq:O10} \\
	\mathcal{O}_{11} &=& i\Schi \cdot \qmN \ , \label{eq:O11} \\
	\mathcal{O}_{12} &=& \Schi \cdot \left( \SN \times \vec{v}^\perp \right) \ , \label{eq:O12} \\
	\mathcal{O}_{13} &=& i \left( \Schi \cdot \vec{v}^\perp \right) \left( \SN \cdot \qmN \right) \ , \label{eq:O13} \\
	\mathcal{O}_{14} &=& i \left( \Schi \cdot \qmN \right) \left( \SN \cdot \vec{v}^\perp \right) \ , \label{eq:O14} \\
	\mathcal{O}_{15} &=& - \left( \Schi \cdot \qmN \right) \left[ \left( \SN \times \vec{v}^\perp \right) \cdot \qmN \right] \ , \label{eq:O15}
\end{eqnarray}
\end{subequations}
up to factors of $q^2$ (which, as a scale-invariant quantity, allows $q^{2n}\mathcal{O}_i$ to be an operator as well). Then, the total Hamiltonian will be
\begin{equation} \label{eq:hamiltonian}
    \mathcal{H} = \sum_{x=n,p} \sum_{i=1}^{15} c_i^x \mathcal{O}_i^x \quad ,
\end{equation}
where we now consider both protons and neutrons as targets and the thirty coupling coefficients $c_i^x$ are free parameters of the interaction theory.

The EFT does not assume any characteristics about the DM particle, other than requiring its interaction with the target be nonrelativistic and mediated by a massive spin-0 or spin-1 particle, e.g., a $U(1)$ gauge boson $A'_\mu$. For this paper, we will consider a spin-$1/2$ weakly interacting massive particle (WIMP) as our dark matter, and we will assume its mass is within the range $10-10^3$\,GeV. In this mass region the DM in our neighborhood of the galactic halo will be nonrelativistic, and any nuclear recoils from the DM will be $\mathcal{O}(100\,\text{keV})$~\cite{jkg1996}.

While the operators Eqs.~\eqref{eq:operators} are constructed assuming Galilean-invariance in the nonrelativistic regime, one can in principle start with a Lorentz-invariant relativistic effective theory and carry out a nonrelativistic reduction. This was done in Ref.~\cite{xia2019pandax} in the EFT framework; see Sec.~VI of Ref.~\cite{Heimsoth2023} for a discussion of the reduction of relativistic couplings to the Galilean EFT operators or Appendix~A. in Ref.~\cite{hoferichter2019} for a comparison of Galilean EFT and an effective theory derived from quark-WIMP interactions.

For our work here we only consider WIMP-nucleon coupling through one-body operators. Previous work~\cite{PhysRevD.86.103511} also considered  the effect of two-body currents derived from chiral effective field theory, albeit approximated as effective one-body operators through normal-ordering the two-body currents with a Fermi gas background~\cite{PhysRevLett.107.062501}. In that work, inclusion of two-body currents reduced the spin-dependent cross-section by up to $55\%$. While such quenching is important for interpreting any experimental results in terms of bare couplings, we emphasize that our goal here is to focus on the uncertainties from nuclear structure, which are independent of any overall quenching factor.

\subsection{Dark Matter Interaction}

Deriving the differential event rate of DM-target interactions from the nuclear model, DM model, and effective theory is nontrivial. A thorough explanation of the process is given in Sec.~II~B-C in Ref.~\cite{Heimsoth2023}; here we give a brief overview for readability of our analysis.

The DM-nucleus interaction cross section can be factorized into nuclear response functions $W_k^{x,x'}$ and WIMP response functions $R_k^{x,x'}$, where $x,x' \in \{p,n\}$ and $k=1,...,8$ runs over the eight allowed combinations of electroweak operators:
\begin{equation} \label{eq:nucresponse}
    W_X^{x,x'} = \sum_{J_T} \langle \Psi_f || X_{J_T}^x || \Psi_i \rangle \langle \Psi_i || X_{J_T}^{x'} || \Psi_f \rangle \,.
\end{equation}
Here $\Psi_i$~($\Psi_f$) is the initial (final) target wave function, $J_T$ is the angular momentum rank of the operator, and $\langle || \cdot || \rangle$ is the usual reduced matrix element. Since we assume the DM to be heavy and nonrelativistic, very little energy will be transferred to the nucleus (as mentioned in the previous subsection). Thus, we can assume the final wave function of the nucleus will be the same as the initial wave function, i.e., the ground state, and we will make this assumption in the rest of this paper. The WIMP response functions, given in Eq.~38 of Ref.~\cite{Fitzpatrick_2013}, group the EFT operators for the corresponding nuclear response functions and depend on the coupling coefficients $c_i^x$ in Eq.~\eqref{eq:hamiltonian}. 

The matrix elements in Eq.~\eqref{eq:nucresponse} are calculated by summing over single particle orbitals in the nuclear model:
\begin{equation} \label{eq:matrixelements}
    \langle \Psi_f || X_{J_T}^x || \Psi_i \rangle = \sum_{a,b} \langle a || X_{J_T}^x || b \rangle \rho_{J_T}^{fi} (a,b) \quad .
\end{equation}
Here, the nuclear one-body density matrices are
\begin{equation} \label{eq:onebodydens}
    \rho_{J_T}^{fi}(a,b) = \frac{1}{\sqrt{2J_T + 1}} \langle \Psi_f || [\hat{c}_a^\dag \otimes \tilde{c}_b]_{J_T} || \Psi_i \rangle \,,
\end{equation}
where $\hat{c}_a^\dag$ and $\tilde{c}_b$ are the fermion creation and time-reversed annihilation operators, respectively~\cite{edmonds1957angular}. The differential cross section is then~\cite{GORTON2023}
\begin{equation} \label{eq:crosssection}
    \dv{\sigma}{\Erec} \left(v^2, \Erec \right) = \frac{2 m_T}{(2 j_T + 1) v^2} \sum_{x,x'=p,n} \sum_{i=1}^8 R_k^{x,x'} W_k^{x,x'}
\end{equation}
where $j_T$ is the spin of the target nucleus (not to be confused with $J_T$, mentioned earlier) and $v$ is the velocity of the DM particle.

The differential event rate depends on the differential cross section, DM velocity distribution $f(\vec{v})$, and detector characteristics (c.f. \cite{LEWIN199687}):
\begin{equation} \label{eq:diffevrate}
    \dv{R}{\Erec} \left( \Erec \right) = \varepsilon(\Erec) N_T n_\chi \int \dv{\sigma}{\Erec} \tilde{f}(\vec{v}) |\vec{v}| \dd^3 \vec{v} \,.
\end{equation}
Here, $\varepsilon(\Erec)$ is the detector efficiency, $N_T$ is the number of target nuclei in the detector, $n_\chi$ is the local DM number density, and $\tilde{f}(\vec{v})$ is the DM velocity distribution boosted into the rest frame of the detector. For the CUORE detector, we take a constant total efficiency $\varepsilon = 0.883$ which is the product of the detector trigger efficiency (99\%), data selection efficiency (90\%), and anti-coincidence selection efficiency (99.2\%) as were the case in Ref.\,\cite{Alduino2017}.

\section{Uncertainty in EFT Coupling Upper Limits in CUORE}
\label{sec:uncerts}

Until a positive detection of dark matter occurs, experiments can at best place upper limits on potential coupling strengths between WIMP particles and ordinary matter. Uncertainties on these upper limits come from the same sources as would any observation, including theoretical uncertainties from the models used in analyses of the data. Nuclear modeling uncertainties are known to be significant, especially in large nuclei that are frequently used in low-background, high-sensitivity detectors, yet until recently this source was neglected or otherwise crudely estimated. We show in this section how the uncertainties in the nuclear shell models affect uncertainties in the determination of the sensitivity of the CUORE detector to different EFT interaction channels. We focus mainly on tellurium, which dominates the nuclear mass of the \TeO crystal in CUORE, although we also briefly discuss the WIMP interaction with oxygen.

The Cryogenic Underground Observatory for Rare Events (CUORE), located at Gran Sasso National Laboratory, is a tonne scale bolometric experiment composed of an array of 988 tellurium dioxide crystals. Each crystal is 0.75 kg and 5 cubic centimeters. The crystal array is cooled to $\sim$15 mK, and has currently collected over 2 tonne-years of $\mathrm{TeO_{2}}$ exposure. CUORE’s main focus is searching for the neutrinoless double beta decay of $^{130}$Te which would have an observable Q-value at $\sim$2527 keV. CUORE uses Neutron Transmutation Doped (NTD) Ge thermistors to detect thermal phonons generated from  energy depositions, including those from radioactive processes, and is calibrated with a $\mathrm{^{232}Th-\,^{60}Co}$ source, achieving a detector resolution of $\sim$7 keV in the region of interest. While CUORE is intended to search for neutrinoless double beta decay, its high sensitivity, low noise, and large exposure make it a suitable detector to search for WIMP like dark matter through nuclear recoil scattering off the detector nuclei.

Before considering each element separately, we review the calculation of the EFT coupling confidence limit and its nuclear model-derived uncertainty. A more thorough explanation can be found in Sec.\,V of Ref.~\cite{Heimsoth2023}. The differential event rate of a proposed DM-target interaction depends on the EFT coupling coefficients introduced in Eq.~\eqref{eq:hamiltonian} through the WIMP response functions $R_k^{x,x'}$. More specifically, these response functions are quadratic in the $c_i^x$; in the case that only one such coupling is nonzero and the coupling is independent of exchanged momentum $q$, the differential event rate is proportional to $(c_i^x)^2$ (by Eqs.~(\ref{eq:crosssection}-\ref{eq:diffevrate})). We assume here the interaction conserves isospin, so that $R_k^{x,x'}=0$ for $x \neq x'$, i.e., only $\text{p} \rightarrow \text{p}$ and $\text{n} \rightarrow \text{n}$ interactions in the target occur, as is the case for a ground state-to-ground state transition.

Assuming a Poissonian distribution for the number of WIMP-target interaction events, an experiment needs a total event rate of at least $\dv*{R_{\text{min}}}{t} = 2.3$~events/ton\,yr for a positive detection of a DM interaction event at a 90\% confidence level\,\cite{pdg_rev2020}. (We neglect the background rate of the detector in our analysis; a nonzero background would increase $\dv*{R_{\text{min}}}{t}$ and consequently increase the EFT coefficient upper limit, as shown below.) Because of the quadratic dependence of the differential event rate on the EFT coefficients, the 90\% upper limit of a given EFT coefficient, written $c_{i,\text{min}}^x$, can be calculated from a single calculation of the total event rate $\dv*{R_0}{t}$ in the \texttt{dmscatter} code of Ref.~\cite{GORTON2023} using a test value $c_{i,0}^x$ for the EFT coupling:
\begin{equation} \label{eq:mincoeff}
    c_{i,\text{min}}^x = \sqrt{\frac{\dv*{R_\text{min}}{t}}{\dv*{R_0}{t}}} c_{i,0}^x \,.
\end{equation}
We calculate the uncertainties on $c_{i,\text{min}}^x$ for tellurium only, as it dominates the coupling upper limits and their uncertainties over oxygen. We also determine the uncertainties from oxygen qualitatively, as will be explained in the following subsections.

\subsection{Tellurium}

The largest target component by mass in CUORE is tellurium. As natural tellurium is composed of approximately 8\% odd-mass isotopes with nonzero ground state spin (namely, $^{123}$Te and $^{125}$Te, each spin-$1/2$), there is sensitivity to all spin-dependent EFT interactions. This is in contrast to oxygen, which is nearly entirely (99.8\%) \oxy, an even-even isotope with ground state spin zero. Thus, tellurium is a relevant target for probing  spin-dependent WIMP interactions in CUORE. (Naturally occurring xenon is nearly half odd-mass isotopes by abundance; the contrast between xenon- and tellurium based experiments could thus shed vital light on the nature of WIMP-nucleus interactions.)

\begin{figure*}
    \centering
    \includegraphics[width=0.95\linewidth]{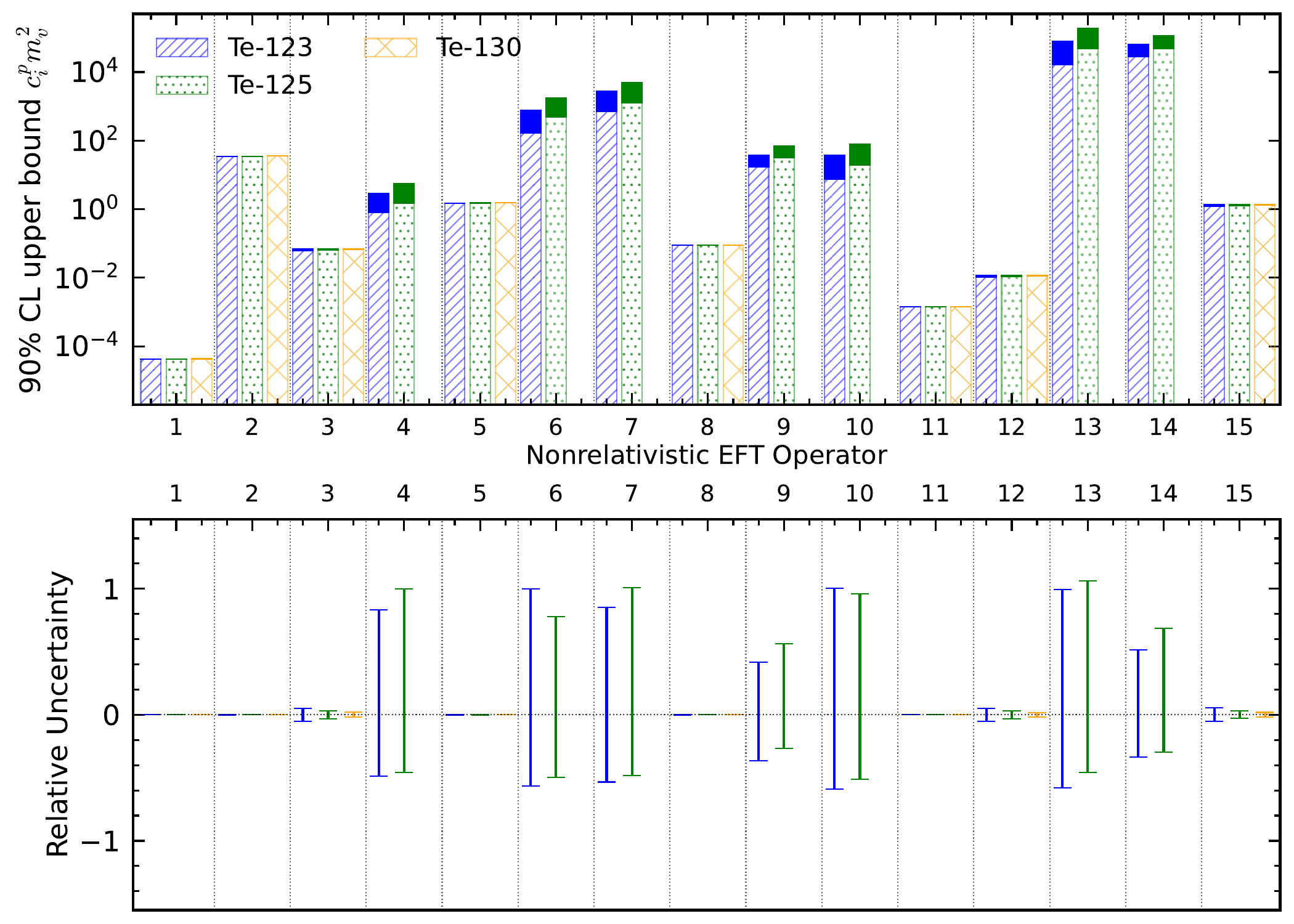}
    \caption{90\% upper confidence limits and uncertainties of Galilean EFT proton couplings for different isotopes of tellurium and WIMP mass $m_\chi = 69.5193$ GeV.}
    \label{fig:te_p_uncerts}
\end{figure*}

To determine uncertainties on the EFT coupling upper confidence limits for WIMP-nucleus interactions with tellurium isotopes, we take the same approach as was performed in Ref.~\cite{Heimsoth2023} for xenon. Nuclear shell-model configuration-interaction calculations of density matrices are costly; to avoid performing many such calculations we started with two different shell model Hamiltonians, GCN~\cite{Caurier:2007wq,Caurier:2010az} and JJ55~\cite{Brown:2005} for tellurium, which are both based on a CD-Bonn potential G-matrix~\cite{HJORTHJENSEN1995125,Machleidt:2000ge} but fit to different experimental data sets. The original papers~\cite{Caurier:2007wq,Caurier:2010az,Brown:2005} found good agreement with tellurium levels and  transition probabilities, and in our more recent work~\cite{Heimsoth2023} we also found good agreement in xenon isotopes. The difference in  observable values is derived from uncertainties in fitting these Hamiltonians to experiment (and, more generally, derived from the difficulty of modeling nuclei).

We constructed a Gaussian ensemble of one-body density matrices  by taking the one-body density matrices from each shell model, GCN and JJ55. We then created a Gaussian probability distribution function (PDF) using the two values for each matrix element to fix the average and the width. Finally, we performed a Gaussian ensemble Monte Carlo (MC) simulation on the one-body nuclear density matrices (Eq.~\eqref{eq:onebodydens}); in each single Monte Carlo run a value for each matrix element is randomly selected  from its associated PDF, and  the resulting randomized one-body density matrix is used in the \texttt{dmscatter} code to calculate a differential WIMP-nucleus event rate (Eq.~\eqref{eq:diffevrate}). A total event rate is calculated by integrating the differential event rate over recoil energy $\Erec$, which we take to be in the range $[1,100]$~keV.

We ran $N=2500$ Monte Carlo iterations for WIMP masses in the range $m_\chi \in [10,1000]$~GeV and nine different tellurium isotopes (the naturally-occurring Te-120, 122, 123, 124, 125, 126, 128, 130 and the unstable Te-129). The best-fit 90\% upper confidence limit on the $c_i^x$ are taken as the median value of the $c_{i,\text{min}}^x$ calculated by Eq.~\eqref{eq:mincoeff} over all MC runs. The uncertainty on the upper limit is calculated using the Feldman-Cousins method~\cite{Feldman:1997qc} to bound 68.23\% ($1\sigma$) of all run values; see Sec.~V and App.~B in Ref.~\cite{Heimsoth2023} for further discussion of this method. Notably, this method produces separate upper and lower uncertainties, which allows the determination of a ``$1\sigma$" interval of a non-Gaussian distribution.

Fig.~\ref{fig:te_p_uncerts} shows the 90\% upper confidence limits of the WIMP-proton EFT coefficients $c_i^p$ and their associated uncertainties for $^{123}$Te, $^{125}$Te, and $^{130}$Te, assuming a WIMP mass $m_\chi = 69.5193$~GeV. At the low end of our WIMP mass region ($\sim$10\,GeV), there is a possibility that the DM particle could excite the nucleus as the DM would have a higher momentum (assuming fixed DM energy density as we do here). This is especially pertinent for $^{125}$Te, which has a low-lying first excited state around $35$\,keV. Since Galilean EFT assumes elastic interactions, we save this scenario for future work. Uncertainties are negligible for couplings that are independent of nucleon spin $\SN$, namely, \Op{1}, \Op{2}, \Op{5}, \Op{8}, and \Op{11}. Operators 3, 12, and 15 all have small ($< 5\%$) uncertainties and contain $\SN \cross \vec{v}^{\perp}$. The remaining operators, being nucleon spin-dependent, do not couple to the even-even isotopes but have substantial uncertainties in the even-odd isotopes, in some cases surpassing $100\%$. These EFT channels also have some of the smallest event rates (and thus the highest EFT upper confidence limits).

\begin{figure*}
    \centering
    \includegraphics[width=0.95\linewidth]{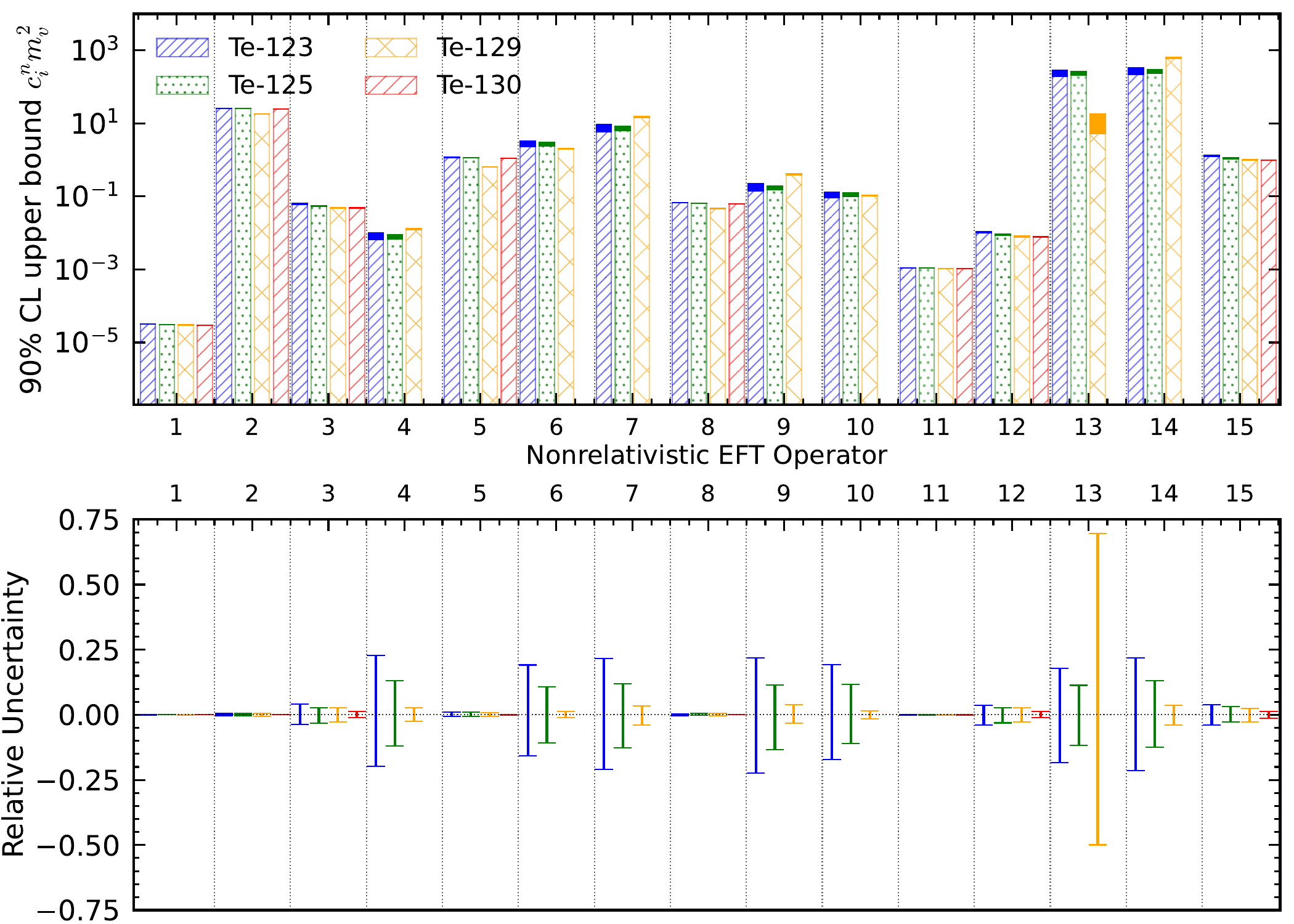}
    \caption{90\% upper confidence limits and uncertainties of Galilean EFT neutron couplings for different isotopes of tellurium and WIMP mass $m_\chi = 69.5193$ GeV. We include $^{129}$Te despite it not being a stable, naturally-produced isotope to highlight both its very large uncertainty in operator 13 and the general trend of decreasing uncertainty with increased nucleon number.}
    \label{fig:te_n_uncerts}
\end{figure*}

Fig.~\ref{fig:te_n_uncerts} provides the analogous 90\% upper confidence limits for the WIMP-neutron couplings at $m_\chi = 69.5193$~GeV, with the addition of $^{129}$Te. The trends in the upper limit uncertainties over the proton-coupled EFT operators are also exhibited by the neutron couplings. Nucleon spin-independent couplings have no significant uncertainty while those depending on $\SN$ do have an appreciable uncertainty, reaching up to 25\% for $^{123}$Te across operators. We include $^{129}$Te in the neutron couplings, despite it being radioactive and only found naturally in trace amounts, to highlight the very large uncertainty of the upper limit of $c_{13}^n$ in this isotope.

\subsection{Oxygen}

%added by CWJ on Oct 22, 2024
%
As the CUORE target is tellurium dioxide, we also computed the WIMP scattering rates off $^{16}$O. We used three different wave functions to generate our reference one-body density matrices. The first and simplest is a single Slater determinant with filled $0s$ and $0p$ shells.  We also computed no-core shell-model (NCSM)~\cite{barrett2013ab} wave functions with two nucleon-nucleon interactions (fitted to low-energy scattering and few-body systems), the Daejeon16 interaction~\cite{shirokov2016n3lo}, as well as the Entem-Machleidt interaction derived from chiral effective field theory at next-to-next-to-next-to-leading-order (N3LO)~\cite{PhysRevC.68.041001}, and evolved by the similarity renormalization group~\cite{PhysRevC.75.061001} to a resolution  $\lambda=2.0 \, \mathrm{fm}^{-1}$; for these calculations we used an $N_\mathrm{max}= 6$ many-body truncation.  

All three wave functions are constructed from harmonic oscillator single-particle states. The only parameter for such harmonic oscillator states is the oscillator length parameter $b = \sqrt{\hbar/M\Omega}$ where $M$ is the nucleon mass and $\Omega$ is the oscillator frequency. For Daejeon16 and the Entem-Machleidt N3LO interactions, we varied $b$ (or, alternately, $\hbar\Omega$) to minimize the ground state energy, at $\hbar \Omega = 17.5$ MeV ($b = 1.540$ fm) for Daejeon16 and $\hbar \Omega = 24$ MeV ($b=1.315$ fm) for Entem-Machleidt N3LO. (These calculations were not fully converged, especially as we left out the three-body forces needed for accurate ground state energies using Entem-Machleidt, but we believe they nonetheless provide a good estimate of the oxygen response relative to tellurium.) 

The oscillator parameter $b$ sets the length scale of the calculations, and it is thus natural to compare the calculated root-mean-square radius against the experimental value of 2.59(7)\,fm~\cite{PhysRevLett.117.052501}.  It is known, however, that NCSM calculations struggle to reproduce experimental radii~\cite{bogner2008convergence,PhysRevC.105.L061302}.  In order to investigate the sensitivity to the length scale, we used different values of $b$ in our calculations. For example, the phenomenological Blomqvist-Molinari formula~\cite{blomqvist1968collective} for $b$ in shell-model calculations would give $b=1.726$ fm. To match the experimental radius, however, 
the simple closed-shell $0s0p$ model required $b=1.764$ fm, while the Daejeon16 and Entem-Machleidt calculations required $b=1.709$ fm and 1.663 fm, respectively, to agree with experiment. 

\begin{figure*}
    \centering
    \includegraphics[width=0.95\linewidth]{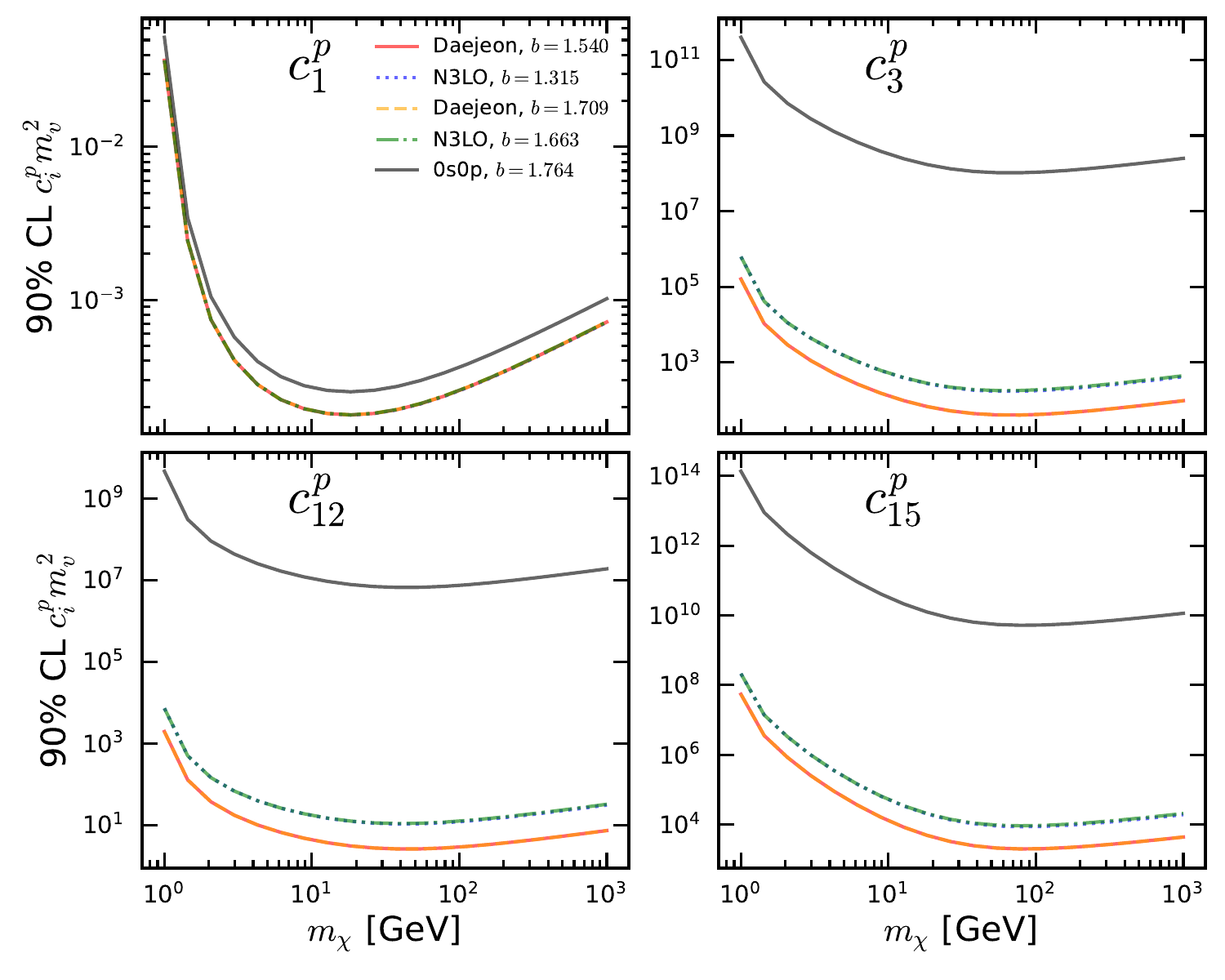}
    \caption{Comparison of 90\% upper confidence limits of Galilean EFT couplings for three nuclear models of \oxy~as a function of WIMP mass $m_\chi$. For two of the  models, denoted Daejeon and N3LO, limits were calculated using two different values for the harmonic oscillator length parameter $b = \sqrt{\hbar/M\Omega}$ where $M$ is the nucleon mass and $\Omega$ is the oscillator frequency, which (among other things) sets the expected radius of the nucleus. The resulting coupling limits are relatively insensitive to the choice of $b$.}
    \label{fig:o16_coeff_limits}
\end{figure*}

Fig.~\ref{fig:o16_coeff_limits} shows the 90\% upper confidence limits in $^{16}$O for each of our three models: the closed-shell $0s0p$ cases, and NCSM wave functions generated using Daejeon16 and Entem-Machleidt N3LO, for four selected EFT couplings. The results shown are typical of the remainder not shown. 

The calculated DM response was not very sensitive to these variations in $b$, even for different operators; a much bigger variation was found between the three models. Because of this lack of sensitivity, we did not tune the value of $b$ for our tellurium results (nor for our previous xenon calculations~\cite{Heimsoth2023}).

As oxygen has nearly an order of magnitude fewer nucleons than tellurium, it is not surprising to find the WIMP upper confidence limits on oxygen is much smaller. Comparing 
against the associated limits from tellurium in Fig.~\ref{fig:te_p_uncerts}, we see the oxygen limits are all at least an order of magnitude higher which implies the event rate $\dv*{R_0}{t}$ (as in Eq.~\eqref{eq:mincoeff}) is almost two orders of magnitude higher for tellurium. Because of the smaller $^{16}$O response, we did not attempt the responses for $^{17,18}$O, which have natural abundances of $\leq 0.2 \%$.

\section{DM Annual Modulation} \label{sec:annualmod}

%\textcolor{purple}{As the Earth orbits around the Sun, its velocity relative to the galactic center changes in magnitude. Assuming a simple, non-rotating galactic dark matter halo, this change in velocity of matter on Earth will affect the center of mass velocity of DM-target interactions. Na{\"i}vely, one expects that as the Earth velocity $\vearth$ (in galactic coordinates) increases, the DM interaction rate increases as more DM particles pass through the Earth in a given amount of time.} Thus, when the Earth is moving in the same direction as the Sun in the galactic frame (which occurs around June), $\vearth$ and the DM interaction rate is maximized; conversely, the interaction rate is minimal in December when the Earth is moving retrograde to the Sun's velocity\,\cite{Freese2013,LEWIN199687}.
Assuming a simple, non-rotating galactic dark matter halo, the motion of the solar system around the galactic center produces a ``dark matter wind'' on Earth. The apparent speed of this wind (and thus the relative kinetic energy that it carries) depends on both the velocity of the solar system as a whole and the velocity of Earth relative to the Sun. When the Earth is moving in the same direction as the Sun in the galactic frame (which occurs around June) the average available energy in a potential DM-target interaction is maximized. This higher energy leads to more interactions occurring above detector thresholds, increasing the measured interaction rate. Conversely, when the Earth is moving retrograde to the Sun's velocity, the interaction rate is minimized\,\cite{Freese2013,LEWIN199687}. Since the Earth takes one year to make a full orbit, the period of any changes in the DM interaction rate should be around one year.

We can model this annual modulation in the differential event rate as a Fourier series\,\cite{Freese2013}:
\begin{equation} \label{eq:evrate_fullmodel}
\begin{split}
    \dv{R}{\Erec} \left(v_{\text{min}}, t, t' \right) = M_0 +& \sum_{n=1}^\infty A_n \cos \left(n \omega (t - t')\right) \\
            +& \sum_{n=1}^\infty B_n \sin \left(n \omega (t-t')\right) \quad ,
\end{split}
\end{equation}
where $\omega = 2\pi / (1 \,\text{year})$ and the $A_n$ and $B_n$ are functions of $v_{\text{min}}$, the minimum WIMP velocity that can produce a recoil energy $\Erec$. Setting $t' = 0$, the dominant terms of the sums are those with period one year, i.e., $n=1$. Rewriting $A_1 \equiv M_1$ and $B_1 \equiv M_2$, Eq.~\eqref{eq:evrate_fullmodel} becomes
\begin{equation} \label{eq:evrate_model}
    \dv{R}{\Erec} \left(\Erec, t \right) = M_0 + M_1 \cos \left( \omega t \right) + M_2 \sin \left( \omega t \right) \, ,
\end{equation}
where we now consider $\dv*{R}{\Erec}$, $M_1$, and $M_2$ as functions of the recoil energy $\Erec$. The total modulation amplitude $\Mtot$ and phase $t_0$ are then calculated from $M_1$ and $M_2$:
\begin{align}
    \Mtot &= \sqrt{M_1^2 + M_2^2} \label{eq:Mtot} \\
    t_0 &= \arctan \left( \frac{M_2}{M_1} \right) \quad . \label{eq:t0}
\end{align}

With the measured annual modulation signature in DAMA/LIBRA persisting over 2.86 ton$\,\times\,$year of exposure\,\cite{damalibra2022}, there has been considerable effort to reproduce the result both in the same detector medium (namely, NaI(Tl) crystal)\,\cite{carlin2024cosine100dataset} and in different materials\,\cite{darkside50_2023}. 
%\st{CUORE, composed of 988 \TeO bolometers for a total mass of 741 kg, is primarily searching for neutrinoless double beta decay ($0\nu\beta\beta$) in $^{130}$Te.} 
Due to CUORE's underground location at LNGS, shielding, and cryostat, the detector has very low backgrounds, allowing the experiment to look for potential direct dark matter interactions down to recoil energies of around 3 keV. In principle, CUORE can search for a dark matter annual modulation in their \TeO crystal as a test of the DAMA/LIBRA result.

To test the response of the differential event rate of DM-nucleus scattering and its nuclear model-induced uncertainty in CUORE to the changing $\vearth$ throughout the year, we performed the Gaussian ensemble uncertainty quantification calculation described in Sec.~\ref{sec:theory} for $\vearth = \{216, 220, 224, 228, 232, 236, 240, 244\}$\,km/s in $^{125}$Te. This isotope has an even-odd nucleus with a nonzero ground state spin, allowing it to couple to all fifteen EFT operators (see Fig.~\ref{fig:te_p_uncerts}). We simulated $n=1000$ Monte Carlo nuclear density matrix inputs and calculated the differential event rate at each $\vearth$ for $\Erec \in [1, 100]$\,keV and a selection of WIMP masses $m_\chi \in [10,1000]$\,GeV. Then, for each Monte Carlo run, we fit the annual modulation model Eq.~\eqref{eq:evrate_model} with fixed $\omega = 2\pi /\text{year}$ and free parameters $\{M_0, M_1, M_2\}$.

We find that changes in nuclear shell model inputs do not affect the parameter ratios $\Tilde{M_1} \equiv M_1/M_0$ and $\Tilde{M_2} \equiv M_2/M_0$. Thus, the phase $t_0$ is insensitive to nuclear uncertainties, as $t_0 = \arctan \left(M_2 / M_1 \right) = \arctan ( \Tilde{M_2} / \Tilde{M_1} )$. We can similarly rewrite $\Mtot$ as
$$ \Mtot = M_0 \sqrt{\Tilde{M}_1^2 + \Tilde{M}_2^2} \quad , $$
so the uncertainty in $\Mtot$ due to nuclear modeling is proportional to the uncertainty in $M_0$, the non-modulating portion of the differential event rate. We find $\log M_0$ decreases linearly with increasing $\Erec$ for all WIMP masses and EFT channels.

\begin{figure*}
    \centering
    \includegraphics[width=0.95\linewidth]{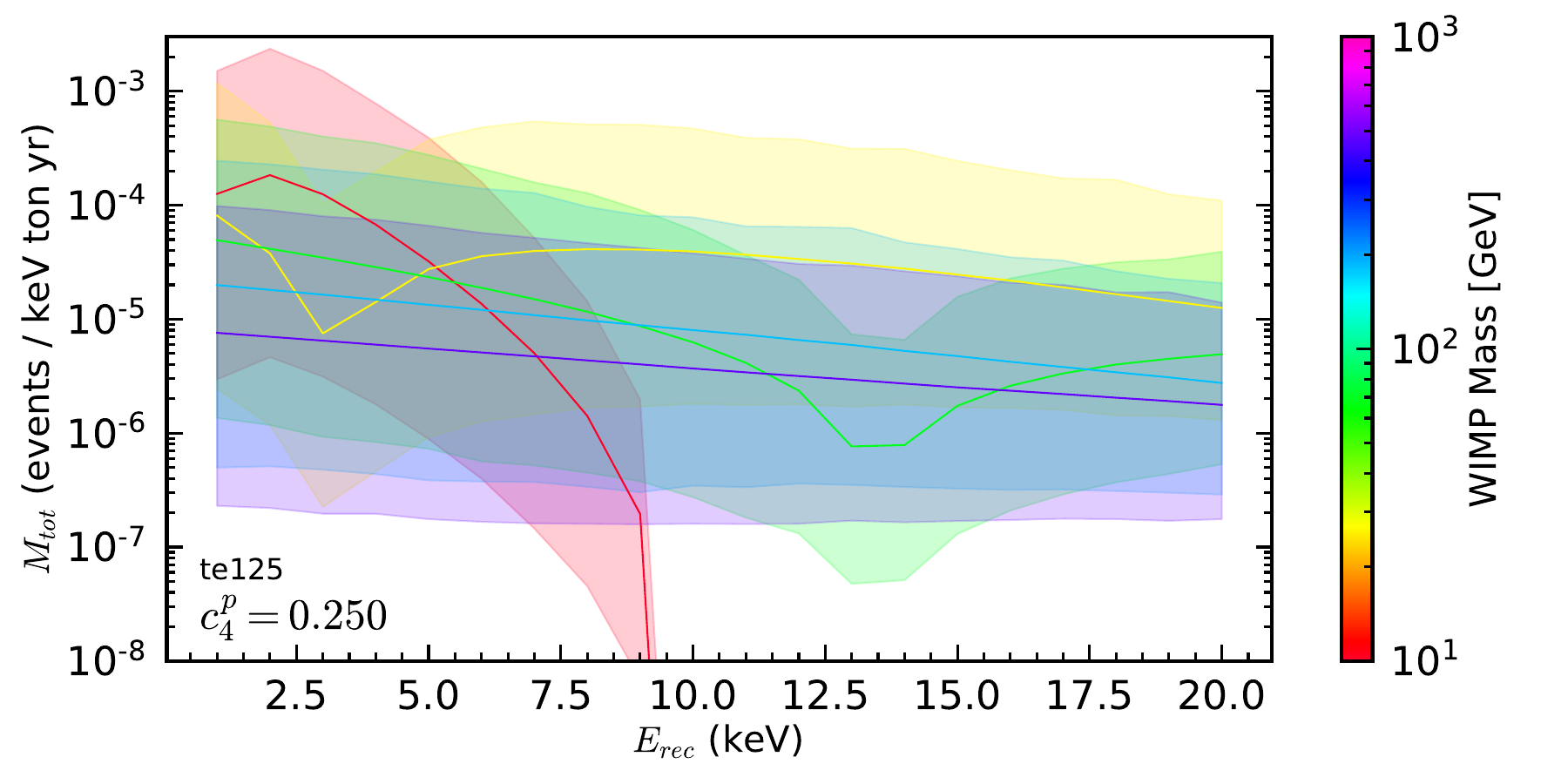}
    \caption{Medians and uncertainties on the dark matter annual modulation amplitude $\Mtot$ as a function for $\Erec$ for the spin-dependent proton coupling $c^p_4$ to Te-125. Multiple $m_\chi$ values are plotted with varying colors.}
    \label{fig:Mtot_uncerts}
\end{figure*}

\begin{figure*}
    \centering
    \includegraphics[width=0.95\linewidth]{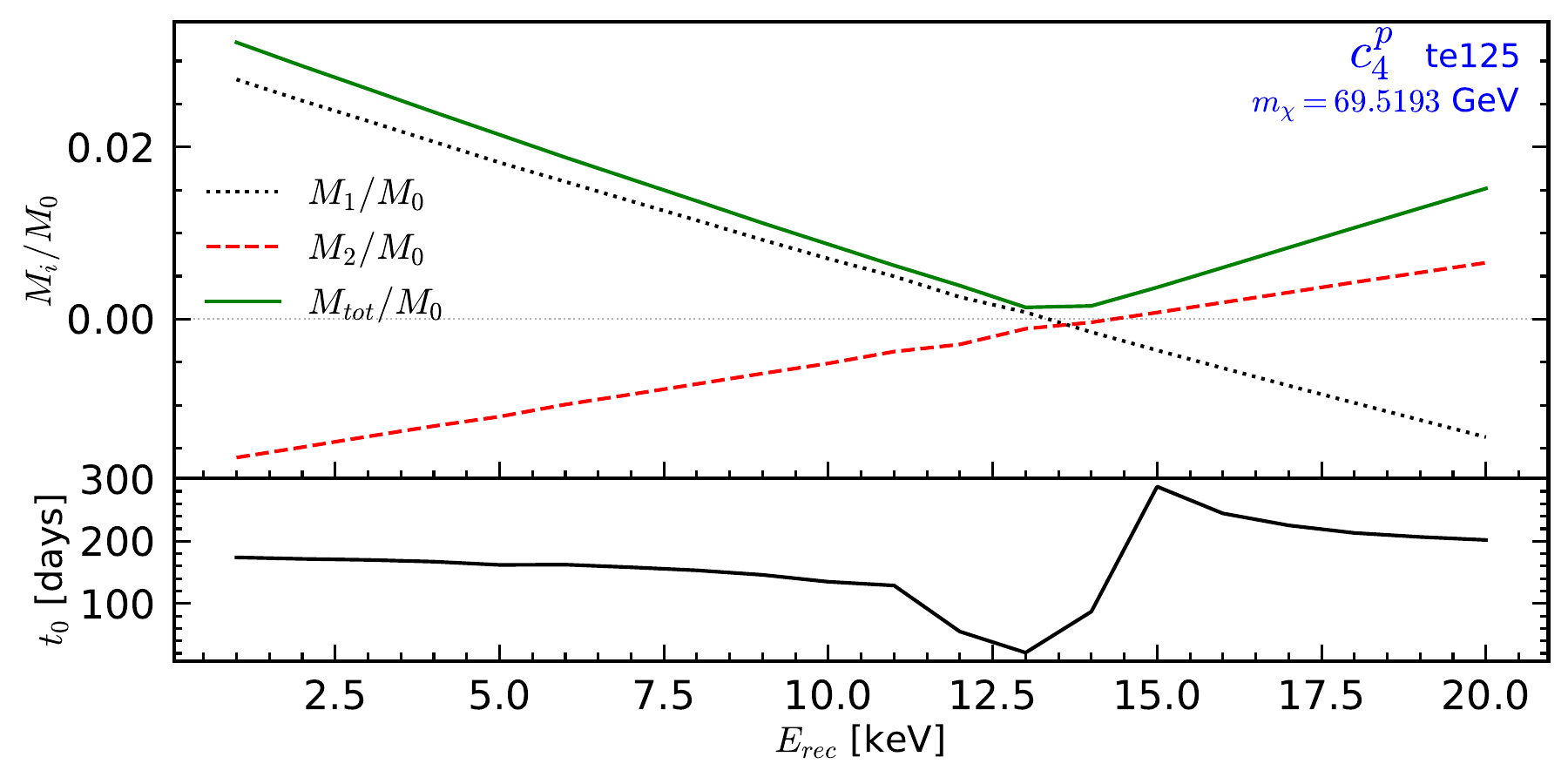}
    \caption{Median annual modulation curve best fit parameters (c.f. Eqs.~\eqref{eq:evrate_model}--\eqref{eq:t0}) as a function of $\Erec$ over $N=1000$ Monte Carlo runs. Note that around $\Erec = 13.5$~keV, $\Mtot$ goes to zero and $t_0$ fluctuates wildly as both $M_1$ and $M_2$ go to zero. Outside of this recoil energy region, the phase is in line with its expected value around 150--200 days, corresponding to a maximum DM interaction rate in the summer months.}
    \label{fig:modulation_values}
\end{figure*}

Fig.~\ref{fig:Mtot_uncerts} shows the median value and uncertainty due to nuclear model uncertainties of the DM annual modulation amplitude $\Mtot$ in $^{125}$Te as a function of $\Erec$ for the proton coupling coefficient of EFT operator $\mathcal{O}_4$, commonly called the spin-dependent coupling. For all $m_{\chi}$ in our range of interest, there is a $\Erec$ wherein $M_{\text{tot}}$ dips down near zero before rebounding, as can be best seen for the yellow and green curves in Fig.~\ref{fig:Mtot_uncerts}. We show an example of this phenomenon for \Op{4} and $m_\chi = 69.5193$~GeV in Fig.~\ref{fig:modulation_values}. At this critical value of $\Erec$, both $M_1$ and $M_2$ flip signs, causing the oscillatory signature to vanish. Thus, in this $\Erec$ regime, the phase is poorly defined and becomes discontinuous in the data before settling back down near the expected value (approx. 150-200 days, corresponding to a maximum event rate in the summer). This phenomenon has been studied before, e.g., in Sec.\,III of Ref.\,\cite{LewisFreese2004} where this crossover recoil energy (denoted $Q_C$) is considered as a probe for WIMP mass.

\section{Discussion}
\label{sec:discussion}

The overall character of the upper limits of EFT couplings and their uncertainties from tellurium in CUORE closely resembles that of xenon in XENON1T, explored in Ref.~\cite{Heimsoth2023}. A quick comparison of Fig.~\ref{fig:te_p_uncerts} above and Fig.~4 in Ref.~\cite{Heimsoth2023} shows strong similarities, from the relative values of the different operator's coupling limits to the significant uncertainties on the spin-dependent couplings that do not interact with even-even isotopes. A few differences stand out, however, such as the much higher relative uncertainties on these spin-dependent couplings in tellurium, with some exceeding $100\%$ upper uncertainty. Further, the coupling limits from tellurium are across all operators approximately ten times higher than the limits set by xenon.  This result suggests that further experimental and theoretical characterization of the nuclear structure of tellurium would be beneficial for improving the accuracy of bounds obtained using tellurium in DM experiments.
%The smaller nuclear model-derived uncertainties on WIMP-xenon couplings also supports this conclusion from the perspective of minimizing systematic uncertainties.

In Fig.~\ref{fig:te_n_uncerts}, we include EFT coupling limits to $^{129}$Te, a radioactive isotope with half-life of approximately 70 minutes, despite no appreciable amount of the isotope existing in natural abundances. The relative uncertainty of \Op{13} for WIMP-neutron coupling in $^{129}$Te is much higher than other isotopes across WIMP masses, while its total event rate (which is inversely proportional to the coupling limit, c.f. Eq.~\eqref{eq:mincoeff}) is also greater. Interestingly, we see this same behavior in the neutron-coupled \Op{13} of $^{131}$Xe, which has the same amount of neutrons ($N = 77$) as $^{129}$Te. This is most likely due to their ground state spin being $3/2$ instead of 0 or $1/2$ for the other even-even and even-odd isotopes in this mass regime, respectively. We do not know, however, why this behavior is only seen in \Op{13}, especially because this operator is a product of two others, \Op{13} = \Op{8}\Op{10}, that do not have significantly larger uncertainties in $^{129}$Te. Further, other operators that contain both nucleon spin $\SN$ and WIMP spin $\Schi$, such as \Op{14}, behave normally. Regardless of the cause of this result, $^{129}$Te is not found in the CUORE detector, so any conclusions about theoretical uncertainties in CUORE dark matter search results will not be affected by this anomaly.

\begin{figure*}
    \centering
    \includegraphics[width=0.95\linewidth]{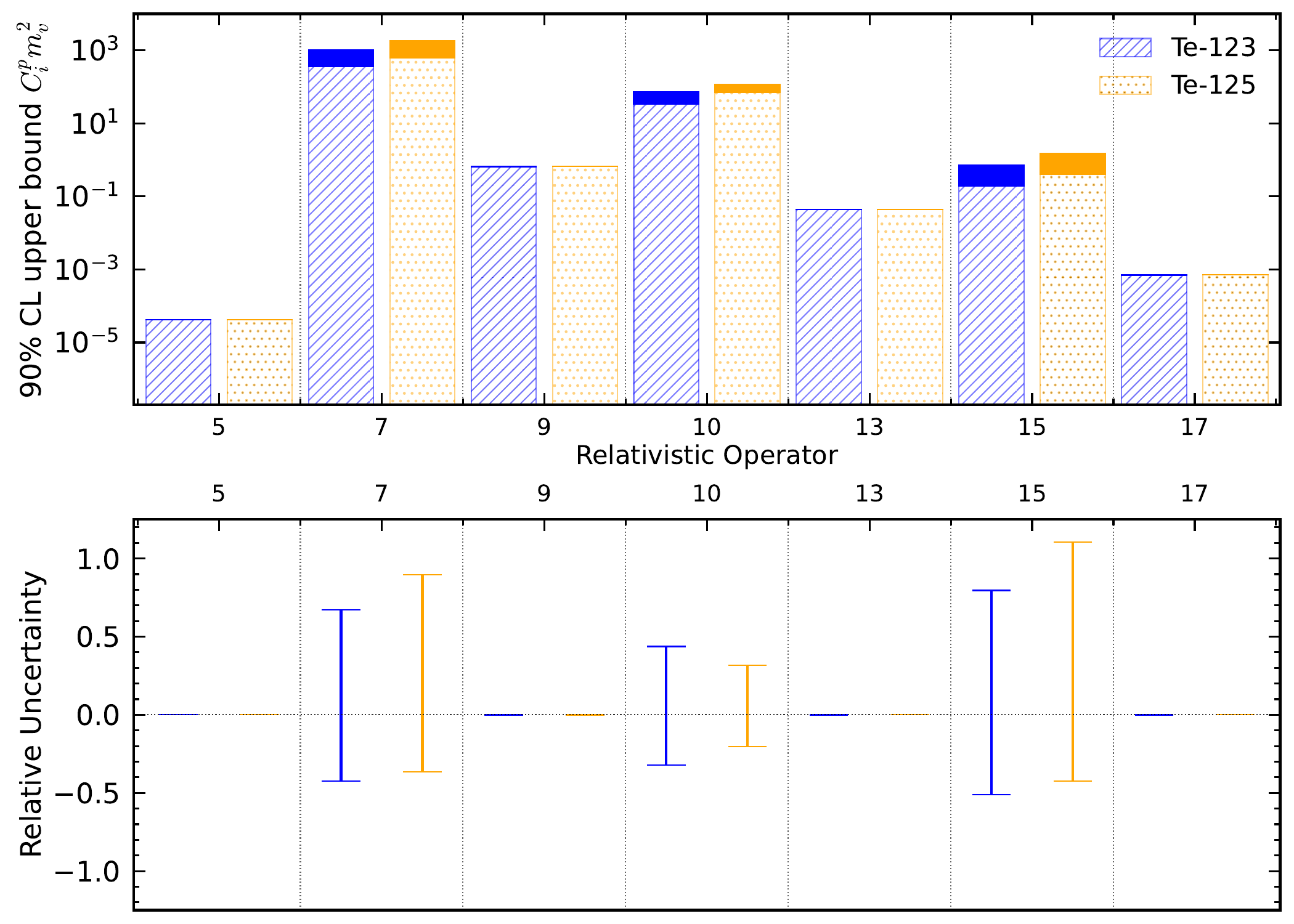}
    \caption{90\% upper confidence limits and uncertainties of relativistic proton couplings for $^{123}$Te and $^{125}$Te and WIMP mass $m_\chi = 69.5193$ GeV. See Ref.~\cite{Heimsoth2023} for definitions of the seven relativistic operators shown here in terms of the Galilean EFT operators. It is no large surprise that the relativistic operators with non-negligible nuclear modeling uncertainties reduce to combinations containing the spin-dependent EFT operators with large uncertainties in Fig.~\ref{fig:te_p_uncerts} in the low momentum exchange limit.}
    \label{fig:te_relp_uncerts}
\end{figure*}

Many of the nonrelativistic operators \Op{i} can be connected to an interaction term between the WIMP and nucleon fields in a relativistic Lagrangian; see Refs.~\cite{Heimsoth2023,xia2019pandax} for the explicit relationships. Simple estimates indicate that the velocities considered in this work are such that non-relativistic theory is applicable, but here we address explicitly whether relativistic effects modify the effects of nuclear model uncertainties on the coefficients of the relativistic Lagrangian interaction terms. Fig.~\ref{fig:te_relp_uncerts} shows the 90\% confidence level upper bounds and their uncertainties for the seven proton-coupled relativistic coefficients $C^{p}_i$ for the two naturally-occurring even-odd isotopes of tellurium, $^{123}$Te and $^{125}$Te. Consistent with our nonrelativistic analysis, the relativistic operators that reduce to spin-dependent Galilean EFT operators (namely, $\mathcal{L}_7$, $\mathcal{L}_{10}$, and $\mathcal{L}_{15}$) have the largest nuclear model-based uncertainties, upward of 100\% for the upper uncertainty of $\mathcal{L}_{15}$ in $^{125}$Te.

We find in Sec.~\ref{sec:annualmod} that the determination of the amplitude, but not the phase, of a DM annual modulation curve is sensitive to nuclear modeling uncertainties. Further, as Fig.~\ref{fig:modulation_values} shows, outside of regions of $(\Erec, m_\chi)$ parameter space where the modulation amplitude is small (e.g., near $\Erec = 12-15$\,keV for EFT coupling $c_4^p$ in Fig.~\ref{fig:modulation_values}), the expected value of $t_0$ is relatively constant. The location of these regions depends on the isotope, although the strength of the reduction in $\Mtot$ does not depend on the nuclear model used (c.f. Fig.~\ref{fig:Mtot_uncerts}), especially for couplings that do not have an appreciable uncertainty due to nuclear modeling uncertainties (i.e., those with negligible uncertainty in Figs.~\ref{fig:te_p_uncerts} or \ref{fig:te_n_uncerts}). 

It is clear from Fig.~\ref{fig:modulation_values} that $\Mtot$ is proportionally larger than the background rate $M_0$ at values of $\Erec$ further away from the low-modulation region. This presents an interesting scenario if the WIMP mass is such that the low-modulation region is within the range of interest for an experiment. The DM modulation signature would be suppressed in some $\Erec$ channels, and determining the position of the dip could set bounds on the value of $m_\chi$. Thus, it is of interest to look for DM modulation over a wide range of $\Erec$ when detector sensitivity permits. 

The ratio $\Mtot / M_0$ was the same for all Monte Carlo runs in the Gaussian ensemble uncertainty quantification for a particular WIMP mass and $\Erec$ in all EFT channels. This implies that nuclear model-derived uncertainties on the modulation amplitude $\Mtot$ are the same as those on the background rate $M_0$. Thus, both experiments that search for modulation signatures and experiments that search for the time-averaged DM interaction rate are sensitive to changes in the nuclear model used in their analyses. 

\section{Conclusions}
\label{sec:conclusion}

We have shown the effect that uncertainties of nuclear modeling can have on dark matter direct detection experiments, in particular on the CUORE detector which uses an array of \TeO crystals to search for nuclear recoils from WIMP-like dark matter interactions. Using the Gaussian ensemble method, we have quantified the uncertainties in determining upper limits on nonrelativistic effective field theory (EFT) coefficients for different isotopes of tellurium in CUORE due to differences in nuclear shell models. We find that the size of the uncertainty varies considerably across the EFT operators, with the largest uncertainties belonging to those operators that depend on nucleon spin $\SN$. We have also considered how these nuclear uncertainties affect WIMP-oxygen interaction rates, finding that different models set different upper limits for EFT couplings across WIMP masses.

We have studied how changes in nuclear modeling affects dark matter annual modulation analyses, finding the modulation amplitude $\Mtot$ can have over 100\% uncertainties for nucleon spin-dependent EFT channels. Moreover, the expected size of $\Mtot$ is dependent on the WIMP mass and nuclear recoil energy, with prominent dips in the amplitude at particular points in the parameter space. We also have found that the ratio $\Mtot / M_0$, where $M_0$ is the background differential event rate, and the modulation phase $t_0$ are independent of changes in the nuclear model. In particular, this shows that uncertainties from nuclear modeling in $\Mtot$ and $M_0$ are identical and thus impact both DM annual modulation searches and background rate experiments equally.

\section{Acknowledgements}

We thank Anna Suliga and Brandon Lem for useful discussions. 
This work was supported in part by U.S. Department of Energy, Office of Science, Office of High Energy Physics, under Award No. DE-SC0019465, and in part by NSF award  PHY-2411495.  DHS would like to gratefully acknowledge the support of the Packard Foundation and Johns Hopkins University.

\bibliography{cuore_dmscatter}

\end{document}